\documentclass{emulateapj}

\usepackage{mathptmx} 
\usepackage{hyperref}

\newcommand{\maxi}{MAXI J0556--332}
\newcommand{\js}{{XTE J1701--462}}
\newcommand{\ks}{KS 1731--260}
\newcommand{\cha}{{\it Chandra}}
\newcommand{\xmm}{{\it XMM-Newton}}
\newcommand{\swift}{{\it Swift}}
\newcommand{\xte}{{\it RXTE}}
\newcommand{\flux}{erg\,s$^{-1}$\,cm$^{-2}$}
\newcommand{\lum}{erg\,s$^{-1}$}
\newcommand{\nh}{$N_{\rm H}$}

\shorttitle{A Hot Neutron Star in \maxi}
\shortauthors{Homan et al.}

\begin{document}

\title{A strongly heated neutron star in the transient Z source \maxi}

\author{Jeroen Homan\altaffilmark{1}, Joel K.\ Fridriksson\altaffilmark{2}, Rudy Wijnands\altaffilmark{2},  Edward M.\ Cackett\altaffilmark{3}, Nathalie Degenaar\altaffilmark{4}, Manuel Linares\altaffilmark{5,6}, Dacheng Lin\altaffilmark{7}, and  Ronald A.\ Remillard\altaffilmark{1} }

\altaffiltext{1}{MIT Kavli Institute for Astrophysics and Space Research, 77 Massachusetts Avenue 37-582D, Cambridge, MA 02139, USA;  jeroen@space.mit.edu}

\altaffiltext{2}{ Anton Pannekoek Institute for Astronomy, University of Amsterdam, Postbus 94249, 1090 GE Amsterdam, The Netherlands}

\altaffiltext{3}{Department of Physics \& Astronomy, Wayne State University, 666 W.\ Hancock St., Detroit, MI 48201, USA}

\altaffiltext{4}{Department of Astronomy, University of Michigan, 500 Church Street, Ann Arbor, MI 48109, USA}

\altaffiltext{5}{Instituto de Astrof{\'i}sica de Canarias, c/ V{\'i}a
  L{\'a}ctea s/n, E-38205 La Laguna, Tenerife, Spain}

\altaffiltext{6}{Universidad de La Laguna, Departamento de
  Astrof{\'i}sica, E-38206 La Laguna, Tenerife, Spain}

\altaffiltext{7}{Space Science Center, University of New Hampshire, Durham, NH 03824, USA}

\begin{abstract}

We present \cha, \xmm, and \swift\ observations of the quiescent neutron star in the transient low-mass X-ray binary \maxi. Observations of the source made during outburst (with the {\it Rossi X-ray Timing Explorer}) reveal tracks in its X-ray color--color and hardness--intensity diagrams that closely resemble those of the neutron-star Z sources, suggesting that \maxi\ had near- or super-Eddington luminosities for a large part of its $\sim$16 month outburst. A comparison of these diagrams with those of other Z sources suggests a source distance of 46$\pm$15 kpc. Fits to the quiescent spectra of \maxi\ with a neutron-star atmosphere model (with or without a power-law component) result in distance estimates of 45$\pm$3 kpc, for a neutron-star radius of 10 km and a mass of 1.4 $M_\odot$. The spectra show the effective surface temperature of the neutron star decreasing monotonically over the first $\sim$500 days of quiescence, except for two observations that were likely affected by enhanced low-level accretion. The temperatures we obtain for the fits that include a power-law ($kT_{\rm eff}^{\infty}$=184--308 eV) are much higher than those seen for any other neutron star heated by accretion, while the inferred cooling ($e$-folding) timescale ($\sim$200 days) is similar to other sources. Fits without a power-law yield higher temperatures ($kT_{\rm eff}^{\infty}$=190--336 eV) and a shorter $e$-folding time ($\sim$160 days). Our results suggest that the heating of the neutron-star crust in \maxi\ was considerably more efficient than for other systems, possibly indicating additional or more efficient shallow heat sources in its crust.

\end{abstract}


\section{Introduction}\label{sec:intro}

In the past $\sim$15 yr the study of cooling neutron stars in transiently accreting X-ray binaries has emerged as a new approach to understanding the structure of neutron stars. \citet{brbiru1998} discussed how the neutron-star crust in transiently accreting systems could be driven out of thermal equilibrium with the neutron-star core.  Some of the heat that is generated by non-equilibrium nuclear reactions deep in the neutron-star crust during an outburst  \citep{hazd1990} is radiated away in quiescence and this should result in observable cooling of the neutron-star surface, especially in systems that have undergone long-duration outbursts \citep{wimima2001,rubibr2002}. By following the detailed evolution of the surface temperature during quiescence one can then extract  information on the  properties of the neutron-star crust.

The neutron-star transient \ks\ provided the first opportunity to test this when it returned to quiescence following an outburst of more than 12.5 yr \citep{wimima2001}. During its first year in quiescence \ks\ indeed showed significant cooling \citep{wiguva2002}, and this continued during the next decade \citep{cawili2006,cabrcu2010}. Following \ks, cooling of neutron stars heated by transient accretion has been studied in a handful of systems: MXB 1659--29 \citep{winomi2003,wihomi2004,cawili2006,cabrcu2013}, \js\ \citep{frhowi2010,frhowi2011}, EXO 0748--676 \citep{dewiwo2009,dewora2011,demecu2014,diboco2011}, and IGR J17480--2446 \citep{debrwi2011,dewibr2013}. These systems had outburst durations ranging from $\sim$11 weeks to $\sim$24 yr, and the outburst-averaged luminosities ranged from a few percent of the Eddington luminosity ($L_{\rm Edd}$) to $\sim$$L_{\rm Edd}$. The (long-term) cooling seen in these systems has been modeled by a variety of cooling curves from increasingly sophisticated thermal evolution codes \citep[see, e.g.,][]{shyaha2007,brcu2009,pare2012,pare2013,tuagpo2013,mecu2014}. Together this has resulted in valuable insights into the thermal conductivity of the neutron-star crust, the location of heat sources in the crust, and, in some cases, the cooling mechanisms at work in the neutron-star core and crust \citep[e.g.,][]{shgumo2014}.

\begin{table*}
\small
\caption{Log of Quiescent Observations Used for Spectral Fitting }
\label{tab:obs}
\begin{center}
\begin{tabular}{cccccc}
\hline
\hline
Obs.\ & Mission & ObsID & Start Time (UT) & Mid Time (MJD)$^{\rm a}$ & Exposure Time$^{\rm b}$ (ks) \\
\hline
1a & \swift & 00032452004 & 2012 May 7 20:24  	& 56054.9215 &  \phn2.4 \\
1b & \swift & 00032452005 & 2012 May 9 04:31   	& 56056.2274 & \phn2.3 \\
1c & \swift & 00032452006 & 2012 May 11 03:12   	& 56058.1759 & \phn2.7 \\
1d & \swift & 00032452007 & 2012 May 13 03:27   	& 56060.2504 &\phn 2.7 \\
2 & \cha & 14225         & 2012 May 21 02:23   		& 56068.1662 & \phn9.2  \\
3 & \cha & 14226         & 2012 May 28 08:59   		& 56075.4444 & \phn9.1 \\
4 & \cha & 14429         & 2012 Jun 5 22:08   		& 56084.0173 & 13.7 \\
5 & \cha & 14433         & 2012 Jun 25 02:22   		& 56103.2258 & 18.2 \\
6 & \cha & 14227         & 2012 July 29 09:32  		& 56137.5281 & 18.2 \\
7 & \xmm\ (MOS1) & 0700380901  & 2012 Aug 17 09:45 & 56156.5735 & 28.6  \\
 & \xmm\ (MOS2) &    \dots & 2012 Aug 17 09:45   & \dots & 28.7 \\
 & \xmm\ (pn) &   \dots  & 2012 Aug 17 10:45   & \dots & 21.5  \\
8 & \xmm\ (MOS1) & 0700381201    & 2012 Sep 16 23:50   & 56187.3194 & 26.2  \\
 & \xmm\ (MOS2) &   \dots  & 2012 Sep 16 23:50   & \dots & 26.8 \\
 & \xmm\ (pn)&   \dots  & 2012 Sep 17 00:51   & \dots & 17.4   \\
9 & \cha & 14434         & 2012 Oct  2 21:22   & 56203.0145 & 17.9 \\
10 & \cha & 14228         & 2013 Feb 20 18:18   & 56343.9210 & 22.7 \\
11 & \xmm\ (MOS1) & 0725220201         & 2013 Sep 13 23:17   & 56549.2302 & 44.9 \\
 & \xmm\ (MOS2) &    \dots      & 2013 Sep 13 23:18   & \dots & 44.5 \\
 & \xmm\ (pn) &         \dots & 2013 Sep 14 00:19   & \dots & 34.8 \\

\hline

\end{tabular}	
\end{center}
{\bf Notes.}\\
$^{\rm a}$ \xmm\ mid times are from EPIC-MOS1 observations.\\
$^{\rm b}$ Live good exposure time after filtering out time intervals affected by particle flaring
\end{table*}

In this paper we present the first observations in quiescence of the cooling neutron star in \maxi, an X-ray transient that was discovered in 2011 January  with {\it MAXI} \citep{manesu2011}. The source had been in outburst for more than 16 months when it returned to quiescence in 2012 May. Although conclusive proof, such as thermonuclear bursts or pulsations, is still missing, there is compelling evidence that the source contains a neutron star. First, the optical-to-X-ray and radio-to-X-ray flux ratios \citep{ruledo2011,cotzco2011} were found to be similar to those of Galactic neutron-star low-mass X-ray binaries (NS-LMXBs). Second, {\it Rossi X-ray Timing Explorer (RXTE)}  data of \maxi\ revealed strong similarities to the so-called ``Z sources'' (\citealt{holiva2011}; see also \citealt{suyama2013}). The Z sources form a class of near- and super-Eddington NS-LMXBs that typically trace out distinct three-branched tracks in their X-ray color--color diagrams (CDs) and hardness--intensity diagrams (HIDs). Combined with other properties (e.g., correlated spectral/timing behavior, scarcity of type-I X-ray bursts) these track shapes set the Z sources apart from the less-luminous ``atoll sources''.
 \maxi\ was found to switch between two types of Z source behavior \citep{holiva2011}, making it very similar to \js\ \citep{lireho2009,hovafr2010}, the first transient Z source, which accreted at near/super-Eddington rates for the majority of its 19 month outburst.   \js\ has shown the hottest neutron-star crust of the cooling systems studied thus far \citep{frhowi2010,frhowi2011}, presumably as the result of its high average mass accretion rate during outburst. A third indication for a neutron-star accretor comes from a dynamical study by \citet{codaca2012}. Assuming a high inclination, inferred from dips in the X-ray light curves, they obtain a compact object mass ($\sim$1.2 $M_\odot$) that is consistent with that of a neutron star. However, we point out that the observed dips are part of the Z source phenomenology \citep{hovafr2010} and do not necessarily indicate a high inclination. Lower viewing angles, and hence higher compact object masses, can therefore not be ruled out.

The fact that \maxi\ had a relatively low peak flux ($\sim$80 mCrab)  suggests that it has a much larger distance than \js\ (which has an estimated distance of 8.8$\pm$1.3 kpc; \citealt{lialho2009}). Based on a comparison of the CD/HID tracks of the two sources, \citet{holiva2011} estimated the distance to  \maxi\ to be  20--35 kpc, which, combined with its  Galactic position ($l$=238$\fdg$9, $b$ = --25$\fdg$2), would place it far out in the Galactic halo. \citet{mamira2011} reported the detection of a strong emission line near 24.8 {\AA} in \xmm\ spectra, which could be identified as the Ly$\alpha$ transition of N {\sc vii}. This could imply a donor with an unusually  high
N/O abundance. Detection of strong N {\sc iii} emission lines in optical outburst spectra, by \citet{codaca2012}, confirmed this. These authors also report candidate optical periods of 16.4 hr and 9.8 hr.

 We started observing \maxi\ with \cha\ and \xmm\ when it returned to quiescence in 2012 May. In this paper we report on the spectral evolution of \maxi\ during  its first $\sim$500 days in quiescence. In Section \ref{sec:obs} we provide a brief description of the observations and our data analysis. The outburst properties, an improved  distance estimate, and fits to the quiescent spectra from the initial cooling phase are presented in Section \ref{sec:results}. The quiescent properties of \maxi\ are discussed in Section \ref{sec:disc}. 

\begin{figure*}[t] 
\epsscale{1.0} 
\plotone{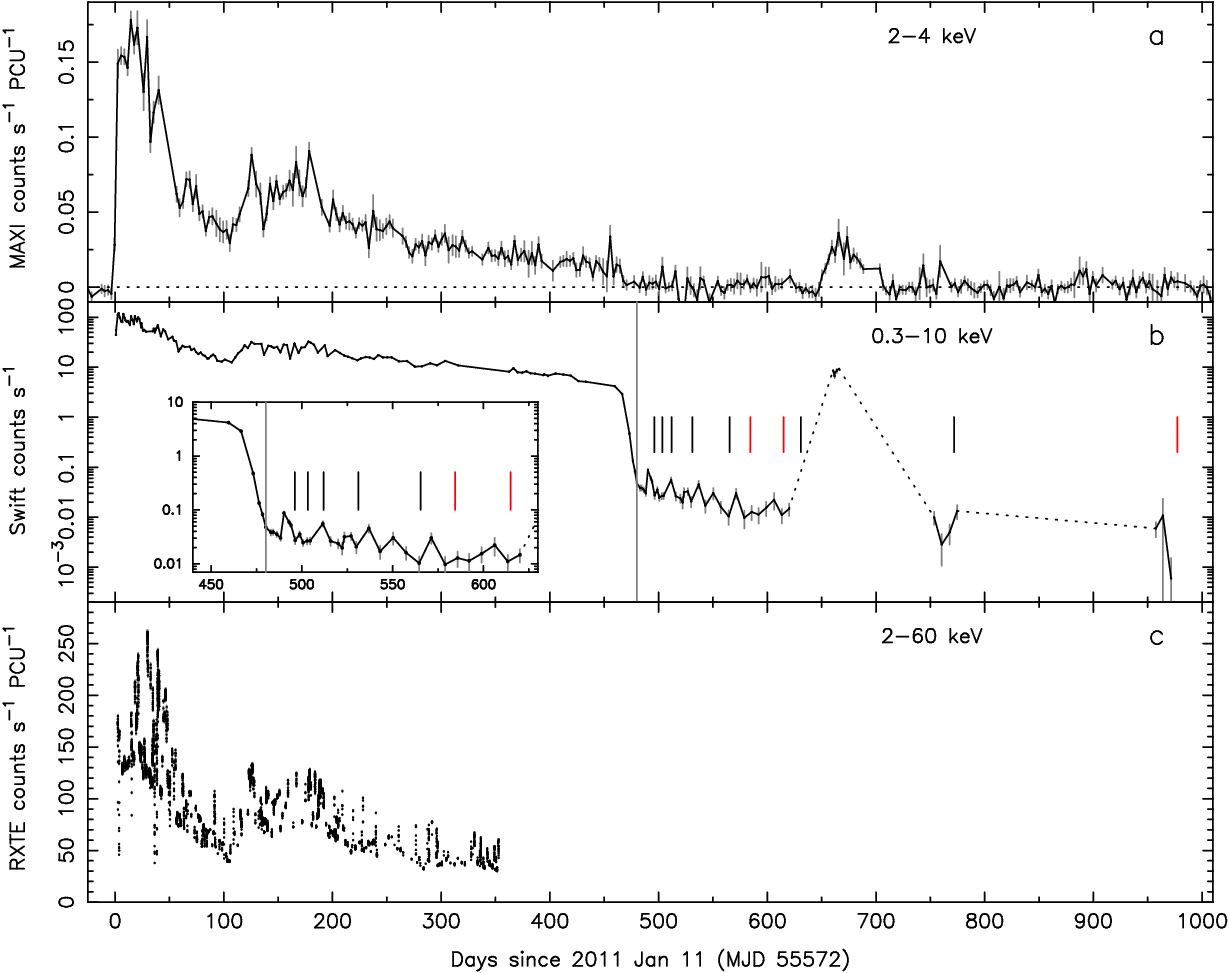} 
\caption{(a) {\it MAXI}/GSC, (b) \swift/XRT,  and  (c) \xte/PCA light curves  of \maxi. Data points represent 3 day averages for the {\it MAXI} light curve, averages over single observations for the \swift\ light curve, and averages over time intervals with lengths between 48 s and a few hundred seconds for the \xte\ light curve. Note that the vertical axes for the {\it MAXI} and \xte\ panels have linear scales, whereas the vertical axis for \swift\ panel is logarithmic (to allow for a more detailed view of the decay and quiescent phases). The  short vertical black and red lines in panel (b) indicate the times of the \cha\ and \xmm\ observations, respectively. The long gray line in panel (b) marks the approximate end of the outburst (MJD 56052.11). The four \swift\ observations listed in Table \ref{tab:obs} correspond to the first four data points after this line. The inset in panel (b) shows an enlarged version of the end phase of the outburst and the start of quiescence.} 
\label{fig:curves}
\end{figure*}

\section{Observations and data analysis}\label{sec:obs}

We analyzed observations of \maxi\ from a variety of X-ray instruments. These observations cover the outburst as well as the quiescent phase of \maxi. The details of the observations are described in Sections \ref{sec:maxi}--\ref{sec:xmm}. A log of the observations specifically used for our study of the quiescent properties of \maxi\ is given in Table \ref{tab:obs}. 

\subsection{{\it MAXI}}\label{sec:maxi}

We used data from {\it MAXI}/Gas Slit Cameras (GSC) \citep{minasu2011} to create a long-term light curve for \maxi. The light curve was constructed from 1 day averages in the 2--4 keV band (which had the highest signal-to-noise ratio). Data points with errors on the count rate larger than 0.025 counts\,s$^{-1}$ were removed and data were further rebinned to 3 day averages to increase the signal-to-noise. The resulting light curve is shown in Figure \ref{fig:curves}a.

\subsection{\swift}\label{sec:swift}

The outburst  and (early) quiescent phases of \maxi\ were monitored densely with the X-Ray Telescope (XRT) onboard \swift\ \citep{buhino2005}. In total 177 XRT observations were made. A long-term 0.3--10 keV light curve was constructed using the online \swift/XRT data products generator\footnote{\url{http://www.swift.ac.uk/user\textunderscore objects/}} \citep{evbepa2009}; it is shown in Figure \ref{fig:curves}b. We performed a spectral analysis of XRT observations that cover the last few weeks of the decay and the first week of quiescence (see Table \ref{tab:obs} for details of the four quiescent observations). 

The XRT observations were made in Windowed Timing and Photon Counting modes and they were analyzed with HEASOFT v6.12. For the Photon Counting mode observations, the source location was determined with {\tt ximage}. Source spectra were extracted from a circle with a 40\arcsec\ radius; background spectra were extracted from an annulus with inner/outer radii of 50\arcsec/80\arcsec, centered on the source. For the Windowed Timing mode observations, source spectra were extracted from a one-dimensional region with a width of 60\arcsec, centered on the pixel with the highest count rate. The background region had a width of 40\arcsec\ and was located at the end of the one-dimensional image farthest from the source. Exposure maps were created with {\tt xrtexpomap} and these were used with {\tt xrtmkarf} to produce ancillary response matrices. We used the same RMF files from the \swift\ calibration database as were selected by the {\tt xrtmkarf} task. The {\tt AREASCAL} keyword in the background spectra was updated to account for vignetting, bad pixels, and hot columns. The spectra were rebinned to a minimum of 20 counts per bin. However, given the low number of source counts  in the Photon Counting mode spectra from the quiescent phase  ($\sim$75 for each of the four \swift\ observations listed in Table \ref{tab:obs}), these spectra were grouped to at least one photon per spectral bin, and we used the C statistic \citep{ca1979}, modified to account for the subtraction of background counts \citep[the so-called W statistic;][]{waleke1979}, when fitting them. The spectra were fitted in the 0.3--10 keV band, although for the quiescent spectra  there typically were no counts above $\sim$6 keV.

\subsection{\xte}\label{sec:xte}

The \xte\ Proportional Counter Array  \citep[PCA;][]{jamara2006} was used to monitor the outburst of \maxi\ in detail as well, until the end of the mission in early 2012 January. A total of 264 pointed observations were made, 262 of which yielded useful data; the other two did not have PCA data. 

We extracted background-corrected light curves in five energy bands from {\tt standard-2} data from all active proportional counter units (PCUs). The count rates were corrected for changes in the PCA response, using long-term light curves of the Crab for each individual PCU, and then normalized to those of PCU 2. The data were adaptively rebinned in time to achieve a minimum of 16,000 counts per bin in the 2--60 keV band, to obtain a more uniform size in the error bars across different values in count rate; data in other energy bands were rebinned accordingly. We also created a CD and color--intensity diagrams, using the same rebinned data. We define  soft color as the net counts in the 4.0--7.3 keV band divided by those in the 2.4--4.0 keV band and  hard color (hardness) as net counts in the 9.8--18.2 keV band divided by those in the 7.3--9.8 keV band. The intensity we use for the color--intensity diagrams is the net count rate per PCU in the 2--60 keV band. An \xte\ outburst light curve is shown in Figure \ref{fig:curves}c, while the corresponding HID is shown in Figure \ref{fig:hid}a.

\subsection{\cha}

\maxi\ was observed seven times with the back-illuminated S3 CCD chip of the Advanced CCD Imaging Spectrometer \citep[ACIS;][]{gabafo2003} onboard \cha.  All observations were made in FAINT mode, with a 1/8 subarray and a frame time of 0.4 s. The data were analyzed using CIAO 4.5 (CALDB 4.5.6) and ACIS Extract version 2012 November 1 \citep{brtofe2010}. As a first step, the {\tt chandra$\_$repro} script was run to reprocess  the data from all the observations. The data were checked for periods of enhanced background; only 0.35 ks from observation 14434 had to be removed. Further analysis was performed with the help of ACIS Extract.

An improved X-ray position was obtained (J2000): $\alpha=05^{\rm h}56^{\rm m}46\fs27$, $\delta=-33\degr10\arcmin26\farcs5$, with an estimated 90\% uncertainty less than $0\farcs3$. This is consistent with the positions of the optical \citep{ha2011} and radio counterparts  \citep{cotzco2011}, as well as  the less accurate \swift/XRT X-ray position reported by \citet{keevkr2011}.

Source spectra
were extracted from near-circular polygon-shaped regions modeled on the
\cha/ACIS point-spread-function (PSF).  The
source extraction regions had a PSF enclosed energy fraction of $\sim$0.97
(for a photon energy of $\sim$1.5 keV) and a radius of $\sim$1$\farcs$9. For all observations except 14227 and 14429 the background was extracted from an annulus with inner radius 5$\farcs$9 and outer radius of 13$\farcs$3--13$\farcs$5. For ObsID 14227 we additionally masked out the readout streak with a 85\arcsec$\times$7\arcsec\ rectangle surrounding the streak, and in this case the outer radius of the background annulus was 14$\farcs$8 (same inner radius as for the other ObsIDs). For ObsID 14429 we masked out the readout streak in the same way, but used an inner/outer radius of 10\arcsec/16$\farcs$6 for the background annulus.  Response files were created using the {\tt mkacisrmf} and {\tt mkarf} tools in CIAO. The spectra were grouped to a minimum signal-to-noise ratio of 4.5 with the ACIS Extract task {\tt ae\_group\_spectrum} and cover the range 0.3--10 keV.

\subsection{\xmm}\label{sec:xmm}

We observed \maxi\ three times with \xmm, using the EPIC-pn \citep{stbrde2001} and EPIC-MOS detectors \citep{tuabar2001}. The detectors were operated in Extended-Full-Frame and Full-Frame modes, respectively, with the thin filter in front of them. The \xmm\ data  were analyzed with the SAS data analysis package, version 12.0.1. The tasks {\tt epproc} and {\tt emproc} were run to produce event files for the pn and MOS observations. All files were inspected for background flares by producing light curves for the entire detectors and only selecting events with {\tt PATTERN=0} (single events) and energies between 10 keV and 12 keV. Data from time intervals with count rates higher than 0.4 counts\,s$^{-1}$ (pn) and 0.15 counts\,s$^{-1}$ (MOS) were discarded. For the first, second, and third \xmm\ observations, $\sim$0\%--1\%, 30\%--40\%, and $\sim$4\%--7\% of the data was removed, respectively. The resulting exposure times are given in Table \ref{tab:obs}. To extract spectra we used {\tt evselect}, with optimal source extraction regions (radii of 20\arcsec--45\arcsec) having been determined by {\tt eregionanalyse}. Background spectra were extracted from rectangular regions (pn) or from circular regions (MOS) on the same chips that the source was located on. The SAS tasks {\tt arfgen} and {\tt rmfgen} were used to create (ancillary) response files. The resulting source spectra were rebinned using the task {\tt specgroup} to obtain a signal-to-noise ratio of 4.5 and the energy range was constrained to 0.3--10 keV.

\begin{figure}[t] 
\epsscale{1.15} 
\plotone{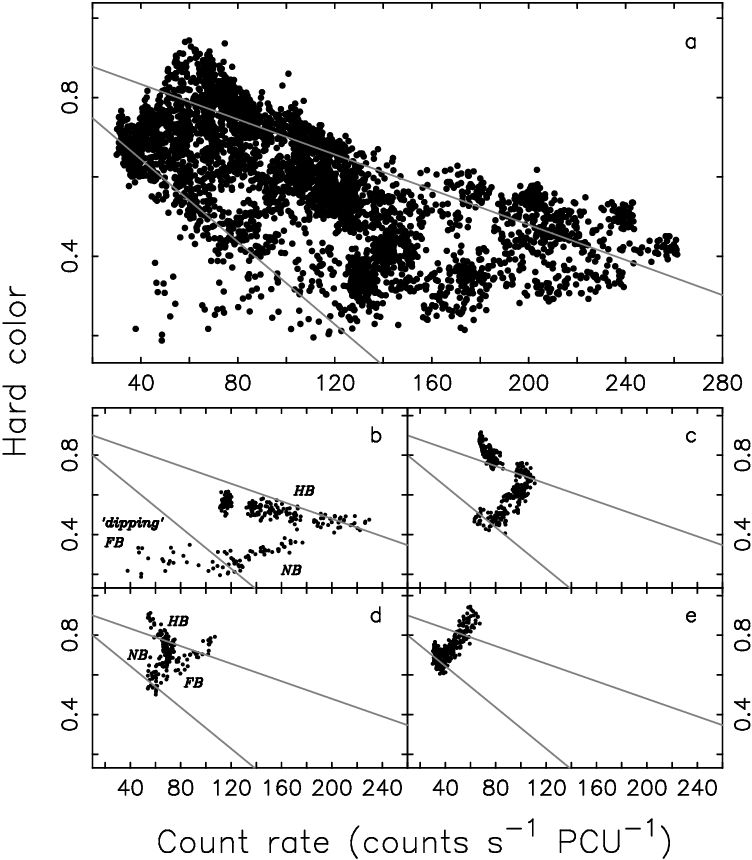} 
\caption{Hardness--intensity diagrams of \maxi.  Panel (a) shows all \xte\ data, while panels (b)--(e) show subsets of the data. Various Z source branches are labeled in panels (b) and (d). The two gray lines represents the approximate paths traced out by the horizontal/normal and normal/flaring-branch vertices as the Z tracks change in shape and position. See Section \ref{sec:xte} for the energy bands used for the count rate and hard color and Table \ref{tab:subsets} for the observations used for the tracks in panels (b)--(e).} 
\label{fig:hid}
\end{figure}

\begin{table*}[t]
\small
\caption{Log of {\it RXTE} Observations Used for Subset Selections in Figure \ref{fig:hid}}
\label{tab:subsets}
\begin{center}
\begin{tabular}{ccccc}
\hline
\hline
Panel & Start Time (MJD) & End Time (MJD) & ObsIDs$^{\rm a}$\\
\hline
b & 55604.72175  &      55610.67675 & 96414-01-03-09 -- 96414-01-04-07 \\
c & 55652.20703  &      55664.05879 & 96414-01-10-00 -- 96414-01-11-04 \\
d & 55704.09332  &      55725.96925 & 96414-01-17-01 -- 96414-01-20-02 \\
e & 55893.82879  &      55924.88101 & 96414-01-44-01 -- 96414-01-48-09 \\
\hline
\end{tabular}	
\end{center}
{\bf Note.}
$^{\rm a}$ First and last ObsIDs (in time) of subset selection.
\end{table*}

\subsection{Spectral Fitting}

There were strong indications that the spectra of observations 4 and 6 (third and fifth \cha\ observations) were affected by residual accretion (see Sections \ref{sec:spectra} and \ref{sec:disc}). For this reason these two observations were fitted together and separately from the main fit. Given the low number of counts in the \swift\ spectra, a different fit statistic was used (see Section \ref{sec:swift}), and observations 1a--d were therefore fitted separately as well. The spectra of the remaining observations (2, 3, 5, and 7--11) were fitted simultaneously. All spectral fits were made with XSPEC v12.8.0 \citep{ar1996}. For the absorption component in our spectral models we used {\tt tbnew} \citep{wialmc2000}\footnote{http://pulsar.sternwarte.uni-erlangen.de/wilms/research/tbabs/}; the {\tt abund} and {\tt xsect} parameters were set to ``wilm" and ``vern" \citep{vefeko1996}, respectively. Note that the use of ``wilm" abundances typically results in \nh\ values that are $\sim$50\% higher than with ``angr'' abundances, as used by  \citet{mamira2011}. 

In addition, we used the pile-up model of \citet{da2001} to correct the \cha\ spectra; pile-up fractions in the \cha\ spectra were as high as $\sim$10\% in the brightest observations. Note that pile-up in the \xmm\ and quiescent \swift\ spectra was negligible ($<$0.1\%). The grade morphing parameter $\alpha$ of the pile-up model could not be constrained very well and was set to 0.6 for all spectra. The actual value of this parameter (0--1) had only minor effects on the resulting fit parameters. For the \xmm\ spectra in the simultaneous fit the frame-time parameter (i.e., the exposure time for each frame) was set to $10^{-6}$ s to disable the pile-up correction. For the highest-flux \cha\ observation (obs.\ 4; 0.15 counts per frame) the temperature shifted upward by $\sim$6\% due to the addition of the pile-up model; for the lowest-flux \cha\ observation (obs.\  10; 0.013 counts per frame) the temperature was not affected significantly. The pile-up model was removed before fluxes were calculated.  No further mention will be made of the pile-up model in Section \ref{sec:results}, where we present our results. 

For the simultaneous fits to the \xmm\ and \cha\ spectra, the parameters for the pn and MOS1/MOS2 spectra were tied for each \xmm\ observation. We further tied the value of \nh\ between all observations and this was done for the distance parameter of the neutron-star atmosphere model and, when applicable, the index $\Gamma$ of the power-law component as well. The best-fit values of these parameters were later used for fits to the spectra of observations 1, 4, and 6, with the exception of the  power-law index, which was allowed to vary in the fits to the spectra of observations 4 and 6. Quoted errors on the fit parameters correspond to 68.3\% confidence limits. The \nh, distance and power-law index were fixed to their best-fit values prior to determining the errors on the temperatures and power-law normalizations. Fixing these parameters allows for more accurate relative temperature measurements, so that the course of the cooling can be followed in more detail, even if the absolute temperatures are not known precisely. We note that  the uncertainty on the absolute temperatures is dominated by our poor knowledge of the distance, with the uncertainties in the \nh\ and power-law index having much smaller contributions.

\section{Results}\label{sec:results}

\subsection{Outburst, Fluence, and Distance Estimate}\label{sec:outburst}

In Figure \ref{fig:curves}b we show the \swift/XRT 0.3--10 keV light curve of \maxi, with each data point representing a single \swift\ observation. It covers the $\sim$16 month outburst, from the end of the rising phase until the source had returned to quiescence. In Figure \ref{fig:curves}c we show the \xte/PCA light curve of \maxi, which covers only part of the outburst. The \xte\ data points represent averages over periods of 48 s to a few hundred seconds. Strong short-term variations (dips and flares) were present during the early stages of the outburst (first $\sim$50 days). During the later stages of the outburst moderate flaring was present. The {\it MAXI} (Figure \ref{fig:curves}a), \swift\ and \xte\ light curves show an initial decline that resulted in a minimum around day 105. This was followed by a rebrightening of the source by a factor of $\sim$2.5. After a broad secondary maximum the source showed a prolonged slow decline, with the \swift/XRT count rate decreasing by a factor of $\sim$4 during a period of $\sim$250 days. This slow decline was followed by a very rapid decay into quiescence, which will be discussed in more detail in Section \ref{sec:decay}.

To estimate the fluence of the outburst we used the \swift/XRT light curve shown in  \ref{fig:curves}b. We integrated the 0.3--10 keV count rate over the outburst, up to day 480 (see gray vertical line in Figure \ref{fig:curves}b), which we define as the end of the outburst (see Section \ref{sec:decay}). Count rates were converted to unabsorbed 0.01--100 keV fluxes using a conversion factor that we derived from fits to \swift/XRT 0.5--10 keV spectra near the peak of the outburst and during the secondary maximum. Fits were made with a model consisting of a black body ({\tt bbody}) and a multi-color disk blackbody ({\tt diskbb}), modified by absorption ({\tt tbnew}, with \nh\ set to 4$\times10^{20}$ atoms\,cm$^{-2}$, a value close to what we obtain from our fits to the quiescent spectra). This model provides good fits to the spectra and results in conversion factors  around 3.5$\times10^{-11}$ \flux\,counts$^{-1}$. While the above spectral model may not be suitable for the harder spectra at low luminosities, for which the conversion factor would likely be larger, the low-luminosity phase of the outburst is short and the effects on our fluence estimate are expected to be very small. Using our conversion factor we obtain a fluence of 2.9$\times10^{-2}$ erg cm$^{-2}$ and an integrated luminosity (or total radiated energy) of 7.1$\times10^{45}(d_{45})^2$ erg, where $d_{45}$ is the distance to the source in units of 45 kpc (see end of this section and Section \ref{sec:spectra} for various distance estimates). The time-averaged luminosity is 1.7$\times10^{38}(d_{45})^2$ \lum, using an outburst duration of $\sim$480 days. This is close to the Eddington luminosity for a 1.4 $M_\odot$ neutron star (1.8$\times10^{38}$ \lum).

\begin{table*}[t]
\small
\caption{Comparison Sources Used for Distance Estimate of \maxi}
\label{tab:distance}
\begin{center}
\begin{tabular}{ccccc}
\hline
\hline
Source & $d$ (kpc) & Method & Ref. & $d_{\rm J0556}$ (kpc)\\
\hline
Sco X-1		& 2.8$\pm$0.3	&	Radio parallax		& \citet{brfoge1999}		& 50--66	\\
LMC X-2			& 50.0$\pm$2.5		&	Location in LMC		& \citet{pigrgi2013} 	& 38--48	\\
XTE J1701--462	& 8.8 $\pm$1.3	&	Type I X-ray burst	& \citet{lialho2009}	& 28--45	\\
%GX 17+2			& 11.9--16.0		&	type I X-ray burst	& \citet{jone2004}	& 60--80	\\
%Cyg X-2			& 11.4--15.3		&	type I X-ray burst	& \citet{jone2004}	& 42--63	\\
%Cir X-1			& 7.8--10.5		&	type I X-ray burst	& \citet{jone2004} &	21--54 \\
\hline
\end{tabular}	
\end{center}
\end{table*}

In Figure \ref{fig:hid}a we show the complete HID of the \xte\ data (which do not cover the entire outburst). The HID reveals a large patch of data points with no clearly identifiable tracks or branches. However, by first inspecting tracks from single observations and later combining data from subsets of consecutive observations, we were able to identify and construct tracks whose shapes are strongly reminiscent of those of the Z sources. In panels b--e of Figure \ref{fig:hid} we show four examples of such relatively complete tracks, in order of decreasing overall count rates. The time ranges covered by these subsets and the corresponding ObsIDs are listed in Table \ref{tab:subsets}. For two of the subset tracks (b and d) we have labeled the Z source branches. %The diagonal lines in Figure \ref{fig:hid} represent the approximate path traced out by the vertex between the normal and (dipping) flaring branch, as the tracks change position and shape in the HID. 
We note that the first and second \xmm\ observations analyzed by \citet{mamira2011} were made during the time intervals used for the  HID tracks shown in panels   \ref{fig:hid}b and  \ref{fig:hid}c, respectively.

The shapes of the HID tracks of \maxi\ gradually evolved as the overall intensity changed. As part of this evolution, the two vertices between the Z source branches showed systematic shifts along paths that could be approximated by straight lines (see gray diagonal lines in Figure \ref{fig:hid}), except during the brief initial rising phase of the outburst (for which we show no separate tracks).  This is similar to what was observed for the HID tracks in the transient Z source \js\ \citep{lireho2009,hovafr2010}. At the highest count rates we observe so-called  Cyg-like Z tracks (b and c), in which the flaring branch is a ``dipping" flaring branch (although the specific shapes of the tracks shown in panels  \ref{fig:hid}b,c are not seen in all Cyg-like Z sources), while at lower count  rates we observe Sco-like Z tracks, with the flaring branch showing higher count rates than the normal/flaring-branch vertex. In the soft-color/intensity diagram (not shown) the Cyg-like tracks show a strong up-turn at the end of the horizontal branch, indications of which can also be seen in the HID track shown in panel  \ref{fig:hid}c. At the lowest count rates (track e) the horizontal branch has disappeared and the normal branch has become very short. All these types of behavior are very similar to what was seen in \js, and strongly suggest that \maxi\ is a transient Z source. 

Quasi-periodic oscillations (QPOs) have also been detected in \maxi\ \citep{bemomu2011}. A preliminary timing analysis performed by \citet{holiva2011} revealed  low-frequency QPOs between $\sim$1.5 Hz and $\sim$35 Hz on the horizontal branch of one of the Cyg-like Z tracks. These QPOs increased in frequency from the horizontal branch upturn towards the horizontal/normal branch vertex, similar to what is seen in other Z sources \citep[see, e.g.,][]{kuvaoo1994,jovaho2002}.

If we assume that similarly shaped Z tracks occur at similar X-ray luminosities in different sources, the HID tracks of \maxi\ can be used to obtain a rough estimate of the distance to the source. To do this, we compare the count rate levels at which certain track shapes are observed in \maxi\ to the count rate levels at which similarly shaped tracks are seen in other Z sources. Specifically, we are using the count rates of the normal/flaring-branch vertex, which is present in all four of the subset HID tracks of \maxi\ shown in Figure \ref{fig:hid}, as a benchmark. The resulting estimates are by no means exact; for example, the effects of interstellar absorption and binary viewing angle are not corrected for, and the method is based on a simple visual comparison. However, the obtained distances should give us an indication of what distance range needs to be explored for our fits to the quiescent spectra. In Table \ref{tab:distance} we list the three Z sources that we use for our distance estimate. We also list the methods that were used to obtain the distances to these three sources. The HID tracks we used for these sources were taken from \citet{fr2011}. Of the four tracks, the one shown in Figure \ref{fig:hid}d is most similar to the tracks of Sco X-1 and LMC X-2, so we used that track for our comparison with those two sources. For \js\ we used the tracks from Figure \ref{fig:hid}b,d,e, since, like \maxi, it showed a variety of track shapes.  From our comparison we find a wide range of distances, 28--66 kpc, and this takes into account the distance uncertainties for the comparison sources. We take the average of the three mid-range values in Table \ref{tab:distance}, 46 kpc, as our best distance estimate and we assume an uncertainty of 15 kpc. While not very constraining, this range rules out the short distance of 1 kpc used by \citet{mamira2011}, which they chose to keep the source well within the Galactic plane. Our distance estimate puts the source well into the Galactic halo, with a distance from the Galactic center of more than 33 kpc and a distance below the plane of more than 12 kpc.

\subsection{Decay and Quiescence}\label{sec:decay}

As can be seen from Figure \ref{fig:curves}b, after a long, slow decline that started around day 200, the decay of \maxi\ started to accelerate around day $460$ of the outburst. Between days 466 and 480 the \swift/XRT count rate dropped by a factor of $\sim$60 with an exponential decay timescale of $\sim$3.3 days. On day 482 the decay had become considerably slower, which we interpret as the source having reached quiescence \citep[see][for similar behavior in \js]{frhowi2010}. The change in decay rate appears to have occurred close to the previous \swift\ observation (on day 480) and we therefore tentatively set the time of the end of the outburst, $t_0$, to be MJD 56052.11 (indicated by the gray vertical line in Figure \ref{fig:curves}b), which is the start time of \swift/XRT observation 00032452003. 

The early quiescent phase of \maxi\ was well covered with \swift; considerable variability can be seen in Figure  \ref{fig:curves}b in the form of small ``flares", lasting several days, on top of a steady, slow decline. During these flares the \swift/XRT count rate typically went up by factors of two to three. Indications of similar variability can still be seen around $\sim$280 days into quiescence. A much stronger reflare started $\sim$170 days into quiescence. During this event the \swift/XRT count rate increased by a factor of more than 600 and from the {\it MAXI} light curve in Figure \ref{fig:curves}a we estimate the duration to have been $\sim$55--60 days. Judging from the {\it MAXI} light curve, the \swift/XRT observations were done around the peak of the reflare; using the same spectral model as in Section \ref{sec:outburst}, we estimate the peak luminosity (0.01--100 keV) to have been $\sim$8.0$\times10^{37}(d_{45})^2$ \lum. Given the sparse \swift\ coverage of the large reflare it is hard to obtain an accurate measurement of its fluence. Using the average {\it MAXI} count rate during the reflare, a  {\it MAXI}-to-\swift/XRT count rate conversion factor of 2.9$\times10^2$, and the same \swift/XRT-to-flux conversion factor as used earlier, we estimate the fluence to be 8.7$\times10^{-3}$  erg\,cm$^{-2}$, which is a factor of $\sim$33 smaller than the fluence of the main outburst. We note that the duration and the time-averaged luminosity  ($\sim$4.4$\times10^{37}(d_{45})^2$ \lum) of the reflare are similar to those of the strongest outbursts of ordinary transient NS-LMXBs, such as Aql X-1 and 4U 1608--52. We also note that the time-averaged luminosity of the reflare is about $\sim$65\% of that of the 11-week outburst of IGR J17480--2446 \citep[$\sim$6$\times10^{37}$ \lum,][]{lialch2012}, during which moderate crustal heating occurred \citep{debrwi2011,dewibr2013}.

\subsection{Quiescent Spectra}\label{sec:spectra}

We started modeling the quiescent spectra of \maxi\ by simultaneously fitting our main set of spectra (i.e.\ those from observations 2, 3, 5, and 7--11) with single-component models. An absorbed power-law ({\tt pileup*tbnew*pegpwrlw}) with $\Gamma$ tied between observations results in a poor fit, $\chi^2_{\rm red}$(degrees of freedom (dof))=2.75 (433), with \nh=(5.1$\pm$0.1)$\times10^{21}$ atoms\,cm$^{-2}$ and $\Gamma=3.17\pm0.03$.     Allowing the power-law index to vary between observations results in a small improvement, $\chi^2_{\rm red}$ (dof)=1.99 (426), with a similar \nh\ and indices between 2.4 and 3.6.

\begin{figure}[b] 
\epsscale{1.1} 
\plotone{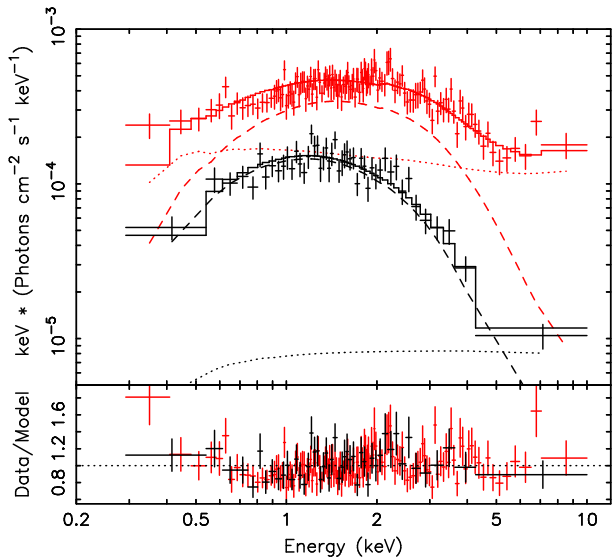} 
\caption{ Top: two unfolded \cha\ spectra of \maxi. The spectrum shown in black is from observation 5, while the spectrum shown in red is from observation 4, which was affected by low-level accretion. Fits with model II are shown as well: the solid, dashed, and dotted lines show the best fit, the {\tt nsa} component, and the {\tt pegpwrlw} component, respectively. Bottom: the corresponding data/model ratios.}  
\label{fig:spectra}
\end{figure}

\begin{table*}
\small
\caption{Spectral Fit Results for Different Distances }
\label{tab:fits}
\begin{center}
\begin{tabular}{cccccccc}
\hline
\hline
Model & $d$ (kpc)                                                   &  \nh\ (10$^{20}$ atoms\,cm$^{-2}$)  &  $\Gamma$                      &  $kT_{\rm eff}^{\infty}$ (eV)$^{\rm a}$  &    $F_{\rm PL}$$^{\rm b}$   & $\chi^2_{\rm red}$ (dof) & $P_\chi$ \\
\hline
{\tt   nsa} 	 	 		& 10$^{\rm c}$ 	                &  35.4$^{+0.6}_{-1.0}$ 	&  $\cdots$	& 105--161	& $\cdots$ 	&  4.34 (434)      & 4$\times10^{-179}$\\
{\tt   nsa} 	 	 		& 20$^{\rm c}$ 	                &  19.2$^{+0.4}_{-0.6}$ 	&  $\cdots$	& 134--218	& $\cdots$ 	&  2.41 (434)      & 2$\times10^{-52}$\\
{\tt   nsa} 	 	 		& 30$^{\rm c}$ 	                &  10.8$^{+0.3}_{-0.6}$ 	&  $\cdots$	& 156--264	& $\cdots$ 	&  1.51 (434)      & 4$\times10^{-11}$\\
{\tt   nsa} 	 	 		& 40$^{\rm c}$ 	                &  5.9$\pm$0.4 	        &  $\cdots$	& 178--306	& $\cdots$ 	&  1.10 (434)      & 8.3$\times10^{-2}$ \\
{\tt  \hspace{0.24cm}  nsa} 	(I)	 		& 46.5$^{+1.0}_{-1.6}$ 	&  3.5$\pm$0.6 	        &  $\cdots$	& 190--329	& $\cdots$ 	&  1.04 (433)      & 0.29 \\
{\tt   nsa} 	 	 		& 50$^{\rm c}$ 	                &  2.3$\pm$0.4        	&  $\cdots$	& 196--342	& $\cdots$ 	&  1.05 (434)      & 0.21 \\
{\tt   nsa} 	 	 		& 60$^{\rm c}$ 	                &  0.0$^{+0.6}_{-0.0}$       &  $\cdots$	& 213--377	& $\cdots$ 	&  1.27 (434)      & 1$\times10^{-4}$\\
\hline
{\tt  nsa + pegpwrlw}	&  10$^{\rm c}$	                &  37.5$\pm$0.5            &   2.58$^{+0.01}_{-0.05}$            & $<$118$^{\rm d}$  & 0.7--15.6   & 2.33 (425)	 & 3$\times10^{-47}$  \\
{\tt  nsa + pegpwrlw}	&  20$^{\rm c}$	                &  18.0$^{+0.7}_{-0.3}$ &   1.48$\pm$0.10                         &  131--195   & 0.22--7.2   & 1.68 (425)	 &  5$\times10^{-17}$ \\
{\tt  nsa + pegpwrlw}	&  30$^{\rm c}$	                &  9.8$\pm$0.3              &   0.94$\pm$0.17                         & 156--249    & 0.13--5.2   & 1.20 (425)	 & 2.5$\times10^{-3}$ \\
{\tt  nsa + pegpwrlw}	&  40$^{\rm c}$	                &  5.6$\pm$0.4              &   0.4$\pm$0.3                             & 177--297    & 0.0--3.7   & 1.00 (425)	 & 0.52 \\
{\tt  nsa + pegpwrlw}	&  43.6$^{+1.0}_{-1.7}$	&  4.3$\pm$0.5              &   0.3$\pm$0.4                             & 185--312    & 0.0--3.3   & 0.98 (424)	 & 0.58 \\
{\tt   \hspace{0.34cm} nsa + pegpwrlw} (II)	&  43.6$^{+0.5}_{-1.5}$	&  4.4$^{+0.7}_{-0.4}$   &   1.0$^{\rm c}$                                    & 184--307    & 0.0--3.2    & 1.00 (425)	 & 0.52 \\
{\tt  nsa + pegpwrlw}	&  50$^{\rm c}$	                &  2.2$\pm$0.4              &   0$^{\rm c}$                                       & 196--337    & 0.0--2.6   & 1.04 (426)	 & 0.31 \\
{\tt  nsa + pegpwrlw}	&  60$^{\rm c}$	                &  0.0$^{+0.6}_{-0.0}$      &   0$^{\rm c}$                                       & 213--374    & 0.0--1.3   & 1.29 (426)	 & 5$\times10^{-5}$ \\
\hline

\end{tabular}	
\end{center}
{\bf Notes.} Observations 1, 4, and 6 were excluded from the fits. All fits used a neutron-star mass $M_{\rm ns}$ of 1.4 $M_\odot$ and a neutron-star radius $R_{\rm ns}$ of 10 km\\
$^{\rm a}$ Temperature range.\\
$^{\rm b}$ Power-law flux range (0.5--10 keV, $10^{-13}$ \flux) .\\
$^{\rm c}$ Parameter was fixed during fit.\\
$^{\rm d}$ The temperatures of several observations peg at the lowest allowed value of $\sim$7 eV. The reported maximum value is for obs.\ 8.
\end{table*}

Next, we tried several absorbed neutron-star atmosphere models.  We only considered hydrogen atmosphere models; this is justified by the presence of strong hydrogen lines in the optical spectra \citep{codaca2012}, which ensures that hydrogen was present in the accreted material. For these models we always assumed a neutron-star mass $M_{\rm ns}$ of 1.4 $M_\odot$ and a neutron-star radius $R_{\rm ns}$ of 10 km, similar to what we used in our previous works. These values were used to convert the obtained unredshifted effective temperatures, $T_{\rm eff}$, to redshifted effective temperatures as measured by an observer at infinity, $T_\mathrm{eff}^\infty=T_\mathrm{eff}/(1+z)$. Here, $1+z=(1-R_\mathrm{S}/R_\mathrm{ns})^{-1/2}$ is the gravitational redshift factor, with $R_\mathrm{S}=2GM_\mathrm{ns}/c^2$ being the Schwarzschild radius. For the mass and radius we used this factor is $\sim$1.306. 

First we performed fits with the neutron-star atmosphere model {\tt nsatmos} of \citet{heryna2006}, {\tt pileup*tbnew*nsatmos}, with the temperatures allowed to vary between observations and the distance left free. This resulted in a poor fit as well, $\chi^2_{\rm red}$ (dof) = 2.32 (433), with the temperatures of observations  2, 3, and 5 pegging at their maximum allowed value, $\log(kT)=6.5$, which is set by the omission of Comptonization in the {\tt nsatmos} model. For this fit we find a distance of 21.6$^{+0.2}_{-0.6}$ kpc. The distance has to be reduced to $\sim$18.8 kpc to avoid the pegging of the temperatures, but this results in a worse fit: $\chi^2_{\rm red}$ (dof) = 2.48 (434).  Leaving the distance free and including a power-law improves the fit ($\chi^2_{\rm red}$ (dof) = 1.37 (424); distance is 27.2$^{+0.8}_{-0.5}$ kpc), but it does not resolve the pegging issue for observations 2, 3, and 5.  We therefore switched from {\tt nsatmos} to the neutron-star atmosphere model of \citet{zapash1996}, {\tt nsa}, which does take into account the effects of Comptonization and allows for higher temperature values. This model,  {\tt  pileup*tbnew*nsa} (hereafter model I), resulted in a greatly improved fit compared to the {\tt nsatmos} model, with $\chi^2_{\rm red}$ (dof) = 1.04 (433) and a $\chi^2$ probability, $P_\chi$, of 0.29. For this fit we obtain \nh=(3.5$\pm$0.6)$\times10^{20}$ atoms\,cm$^{-2}$, a distance of 46.5$^{+1.0}_{-1.6}$ kpc, and a  $T_\mathrm{eff}^\infty$-range of 190--329 eV.  

Adding a power-law component (with indices tied and all normalizations allowed to vary) to the model, {\tt  pileup*tbnew*(nsa+pegpwrlw)}, as was necessary to obtain good fits for the quiescent spectra of \js\ \citep{frhowi2010}, leads to $\chi^2_{\rm red}$ (dof) = 0.99 (424) and $P_\chi$= 0.58. To test the significance of this small improvement we performed a posterior predictive $p$-value (ppp) test \citep{prvaco2002,huvaos2008,cabrcu2013}. We simulated 1000 sets of spectra, based on the best fit with model I. These sets of spectra were then fit with model I and with the above model that included the power-law component.  For none of the 1000 simulated sets of spectra, the latter model yielded an improvement in $\chi^2$ as large as for the observed set of spectra. This suggest that the minor improvement in  $\chi^2_{\rm red}$, as the result of adding a power-law, is statistically highly significant. However, the obtained power-law index of 0.3$\pm$0.4 is extremely low and likely unphysical. Fixing the power-law index to a more reasonable value of 1.0 results in $\chi^2_{\rm red}$ (dof) = 1.00 (425) and $P_\chi$= 0.52; a ppp test indicates that this is still a significant improvement with respect to the fits with model I, with again none of the 1000 simulated sets of spectra yielding a  $\chi^2$ improvement as large as the actual set of spectra. We refer to the model with $\Gamma$ fixed to a value of 1.0 as model II. For this model we obtain \nh=(4.4$^{+0.7}_{-0.4})\times10^{20}$ atoms\,cm$^{-2}$, a distance of 43.6$^{+0.5}_{-1.5}$ kpc, and a  $T_\mathrm{eff}^\infty$ range of 184--307 eV. We note that fits with power-law indices fixed to values closer to those measured in other quiescent NS-LMXBs \citep[e.g.,][]{wihepo2005,frhowi2010} no longer result in improvements compared to model I (i.e., without a power-law component); fits with $\Gamma=1.5$ and $\Gamma=2.0$  give $\chi^2_{\rm red}$ (dof)= 1.04 (425) and $\chi^2_{\rm red}$ (dof)= 1.05 (425), respectively.

\begin{figure*}[t] 
\epsscale{0.7} 
\plotone{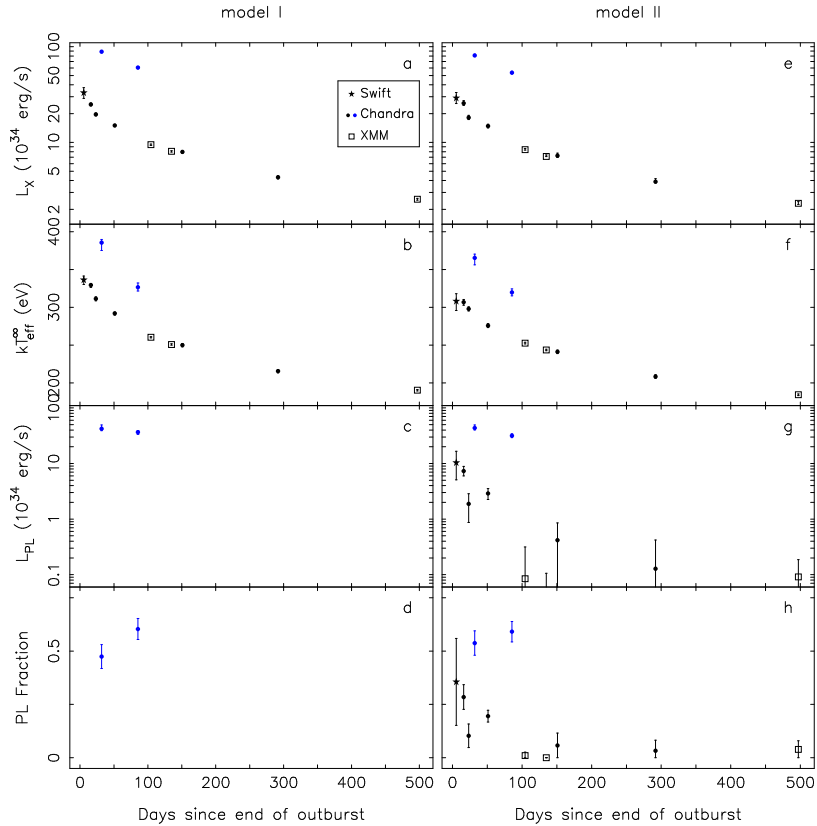} 
\caption{Evolution of the spectral properties of the quiescent neutron star in \maxi\ for two spectral models; the left panels are for a fit with model I (see Table \ref{tab:bestfit1}) and the right panels for a fit with model II (see Table \ref{tab:bestfit2}). From top to bottom we show the 0.5--10 keV luminosity, the effective surface temperature, the 0.5--10 keV power-law luminosity, and the power-law fraction. The blue data points represent the two flare observations that were likely affected by enhanced accretion. The large reflare discussed in Section \ref{sec:decay} occurred between $\sim$170 and $\sim$230 days into quiescence.} 
\label{fig:cool}
\end{figure*}

The \nh\ values obtained with models I and II are only slightly higher than the \nh\ values in the direction of \maxi\ obtained from the H\,{\sc i} surveys of \citet{dilo1990} and \citet{kabuha2005} ($\sim$2.6$\times10^{20}$ atoms\,cm$^{-2}$ and $\sim$2.9$\times10^{20}$ atoms\,cm$^{-2}$) and fall in the range obtained by \citet{mamira2011} from fits to \xmm/RGS spectra, ($\sim$2.1--4.6$\times10^{20}$ atoms\,cm$^{-2}$, using {\tt angr} abundances). In Table \ref{tab:fits} we show how \nh\ and other fit parameters vary as we fix the distance to several different values.  For this we used {\tt nsa} models with and without a power-law component; for comparison, results from models I and II are shown as well. As can be seen, there is a strong dependence of most parameters on the distance; as the distance increases, the column density decreases, the temperatures increase, and (for model II) the power-law component becomes less steep and its flux decreases. The quality of the fits becomes very poor when moving away from the best-fit distances of models I and II.

\begin{table}[t]
\small
\caption{Detailed Spectral Fit Results for Model I ({\tt  pileup*tbnew*nsa})}
\label{tab:bestfit1}
\begin{center}
\begin{tabular}{ccccc}
\hline
\hline
Obs.  & Days Since $t_0$ & $kT_{\rm eff}^{\infty}$ (eV) & $F_{\rm Tot}$$^{\rm a}$ & $F_{\rm PL}$$^{\rm a}$ \\
\hline
1    &  \phn \phn5.5        &   336$\pm$6				&  12.7$\pm$1.6	& $\cdots$ \\
2    &   \phn16.1	&   329$\pm$3				&  9.6$\pm$0.3	& $\cdots$ \\
3    &  \phn 23.3	&   311$\pm$3				&  7.6$\pm$0.3	& $\cdots$ \\
\hspace{0.15cm}4$^{\rm b}$    &   \phn31.9	&   386$^{+4}_{-10}$		&  34.4$\pm$0.9	& 16.3$^{+2.7}_{-1.3}$ \\
\hspace{0.15cm}5$^{\rm b}$   &   \phn51.1	&   292$\pm$2				&  5.81$\pm$0.15	& $\cdots$ \\
6    &   \phn85.4 	&   327$\pm$5 			&  23.4$\pm$0.6	& 14.1$\pm$1.1 \\
7    &   104.5	&   260.4$\pm$0.9	 			&  3.66$\pm$0.05	& $\cdots$ \\
8    &   134.9	&   250.7$\pm$1.0				&  3.12$\pm$0.05	& $\cdots$ \\
9    &   150.9	&   250$\pm$2				&  3.08$\pm$0.10	& $\cdots$ \\
10  &   291.8	&   216$\pm$2				&  1.67$\pm$0.06	& $\cdots$ \\
11  &   497.1    &   190.3$\pm$1.0				&  0.99$\pm$0.02	& $\cdots$ \\
\hline

\end{tabular}	
\end{center}
{\bf Notes.}\\
$^{\rm a}$ Unabsorbed 0.5--10 keV flux ($10^{-13}$ \flux).\\
$^{\rm b}$ For these observations an additional power law was added to the fit model ({\tt  pileup*tbnew*(nsa+pegpwrlw)})
\end{table}

While fits with model I yield acceptable fits for our main set of spectra (and the spectra of observation 1), this is not the case for the two flare observations (obs.\ 4 and 6). Fitting the spectra of these two observations simultaneously with model I and keeping \nh\ and distance fixed to the values obtained from our main set of spectra, we obtain $\chi^2_{\rm red}$ (dof) = 2.06 (248). Adding a power-law with the indices tied leads to a large improvement, $\chi^2_{\rm red}$ (dof) = 1.11 (245) and $P_\chi$= 0.10, with $\Gamma=1.34\pm0.05$. For fits with model II we obtain $\Gamma=1.26\pm0.05$ for these two observations ($\chi^2_{\rm red}$ (dof) = 1.17 (245)). In Figure \ref{fig:spectra} we show example fits to a flare (obs.\ 4) and non-flare spectrum (obs.\ 5) with model II. As can be seen, the relative contribution of the power-law component is much larger in the flare observation.

The full fit results of models I and II, including those of observations 1, 4, and 6,  are reported in Tables \ref{tab:bestfit1} and \ref{tab:bestfit2}, respectively. The $\chi^2_{\rm red}$ (dof) for the fits to the spectra of observation 1 were 1.03 (262) for model I and  0.99 (261) for model II. The evolution of the unabsorbed 0.5--10 keV flux, the neutron-star temperature, the 0.5--10 keV power-law flux, and fractional power-law  contribution in the 0.5--10 keV band are shown in Figure \ref{fig:cool}. As can be seen from Table \ref{tab:bestfit1}  and \ref{tab:bestfit2} and Figure \ref{fig:cool}, there is an overall monotonic decay of the 0.5--10 keV luminosity and the temperature that is interrupted during observations 4 and 6. In those two observations there is a clear increase in both the thermal and power-law fluxes. As we discuss in Section \ref{sec:disc}, these two observations were likely affected by increases in the quiescent accretion rate. The fractional contribution of the power-law component as obtained from model II was quite high in the first few observations, with values in the range $\sim$10\%--50\%, but after day $\sim$100 it has remained below 10\%. The power-law and {\tt nsa} fluxes are strongly correlated for model II, as can be seen from the values in Table \ref{tab:bestfit2}.

\begin{table}[t]
\small
\caption{Detailed Spectral Fit Results for Model II ({\tt  pileup*tbnew*(nsa+pegpwrlw)})}
\label{tab:bestfit2}
\begin{center}
\begin{tabular}{ccccc}
\hline
\hline
Obs.  & Days Since $t_0$ & $kT_{\rm eff}^{\infty}$ (eV) & $F_{\rm Tot}$$^{\rm a}$ & $F_{\rm PL}$$^{\rm a}$ \\
\hline
1    &   \phn\phn5.5        &   308$\pm$11		&  12.7$\pm$0.2			&  5$\pm$3 		\\
2    &   \phn16.1	&   307$\pm$4			&  11.3$\pm$0.6			&  3.2$\pm$0.6	\\
3    &   \phn23.3	&   298$\pm$3			&  8.0$\pm$0.4			&  0.8$\pm$0.4	\\
4    &   \phn31.9	&   365$^{+5}_{-9}$		&  35.7$\pm$1.0			&  19.2$^{+2.5}_{-1.5}$	\\
5    &   \phn51.1	&   276$\pm$3			&  6.5$\pm$0.3			&  1.3$\pm$0.3	\\
6    &   \phn85.4 	&   320$\pm$5			&  23.6$\pm$0.7			&  13.9$\pm$1.1	\\
7    &   104.5	&   252.6$\pm$1.0			&  3.70$^{+0.11}_{-0.03}$	&  0.04$^{+0.10}_{-0.04}$	\\	
8    &   134.9	&   243.6$\pm$1.0			&  3.14$^{+0.11}_{-0.01}$	&  0.0$^{+0.05}_{-0.00}$	\\
9    &   150.9	&   241$\pm$2			&  3.2$\pm$0.2			&  0.2$\pm$0.2	\\
10  &   291.8	&   208$\pm$2			&  1.71$^{+0.13}_{-0.05}$	&  0.06$^{+0.12}_{-0.06}$	\\
11  &   497.1    &   184.3$\pm$1.1			&  1.02$\pm$0.03			&  0.04$\pm$0.04	\\
\hline

\end{tabular}	
\end{center}
{\bf Note.}
$^{\rm a}$ Unabsorbed 0.5--10 keV flux ($10^{-13}$ \flux).
\end{table}

In Figure \ref{fig:sources} we compare the temperature evolution seen in \maxi\ with that of other cooling-neutron-star systems. We also show fits to the data with an exponential decay to a constant. We stress that these are simply phenomenological fits to extract some basic characteristics for a comparison of the different cooling curves; fits with theoretical models are beyond the scope of the current paper.  Fit values are given in Table \ref{tab:expo} and were taken from the literature (see references in table); for \maxi\ we performed fits to the temperatures obtained with models I and II  (excluding observations 4 and 6). The time scales measured for \maxi\ are among the shortest seen, while the temperature drop it has shown is by far the largest of the sources in our sample. There are no clear indications in the cooling curve of \maxi\ that the source is close to reaching a plateau, so the high baseline level from the fit should be treated more as an upper limit.  To investigate the effect of a change in distance on the fit parameters, we also fitted the temperature data from  spectral fits with the {\tt nsa} model and the distance fixed to 20 kpc. For this case we find an exponential decay time scale $\tau$ of 180$\pm$7 days, decay amplitude $A$ of 92$.0\pm$1.2 eV, and constant level $B$ of 129.3$\pm$1.1 eV. Compared to the model I cooling curve the decay time scale is slightly longer, while the normalization and constant level are both significantly lower.

%\begin{figure} 
%\epsscale{1.0} 
%\plotone{f5_power-law.eps} 
%\caption{The relation between thermal ({\tt nsa}) flux and power-law flux, for model II. Fluxes are in the 0.5--10 keV band. Data for observations 4 and 6 are shown in red.} 
%\label{fig:powerlaw}
%\end{figure}

\section{Discussion}\label{sec:disc}

We have studied the color and spectral evolution of \maxi\ during outburst and in quiescence, with \swift, \xte, \cha, and \xmm. During outburst the source showed CD/HID tracks similar to those seen in the neutron-star Z sources, indicating that \maxi\ was radiating close to or above the Eddington luminosity for a large part of the outburst. A comparison of the CD/HID tracks with those of three other Z sources suggests a distance of $\sim$46$\pm$15 kpc. The decay of the outburst of \maxi\ was covered in great detail with \swift\ and allowed us to resolve the  transition from a fast to slow decay, which we identify as the start of quiescence, to within a few days. Our first quiescent observations started within 3 days of the end of the outburst. The quiescent spectra can be modeled well with a neutron-star atmosphere model, with or without a power-law component. We find that the temperature of the neutron-star component decreases monotonically with time, strongly suggesting that the neutron-star crust in this system is cooling, whereas the power-law component (when used) varies more erratically (albeit with an overall decrease as well). On several occasions during early quiescence increases in the flux by factors of up to $\sim$5 were seen. We argue below that these events were likely the results of enhanced low-level accretion in \maxi, similar to what has been seen in other systems. 

\begin{table*}[t]
\small
\caption{Fits to Cooling Curves with an Exponential Decay to a Constant$^{\rm a}$}
\label{tab:expo}
\begin{center}
\begin{tabular}{lcccl}
\hline
\hline
Source  & $\tau$ (days)  & $A$ (eV) & $B$ (eV) & Data References \\
\hline

MAXI J0556--332    &  161$\pm$5                   &  151$\pm$2            & 184.5$\pm$1.5  &  This work (model I)\\
                                &  197$\pm$10                   &  137$\pm$2            & 174$\pm$2 &  This work (model II)\\
                                
IGR J17480--2446  &   157$\pm$62    &  21.6$\pm$4             & 84.3$\pm$1.4 &  \citet{dewibr2013}\\
EXO 0748--676      &  172$\pm$52        &  18$\pm$3       & 114.4$\pm$1.2  & \citet{demecu2014}  \\
XTE J1701--462    &  230$\pm$46   &  35.8$\pm$1.4       & 121.9$\pm$1.5   &  \citet{frhowi2011} \\
KS 1731--260         &  418$\pm$70      &   39.8$\pm$2.3           & 67.7$\pm$1.3     & \citet{cabrcu2010} \\
MXB 1659--29        &  465$\pm$25                  &   73$\pm$2      & 54$\pm$2     &  \citet{cawimi2008}\\
\hline

\end{tabular}	
\end{center}
{\bf Note.}
$^{\rm a}$ $kT_{\rm eff}^{\infty}(t) = A \times e ^ {-t/\tau} + B $, where $t$ is time since the end of the outburst in days.\\
%$^{\rm c}$ parameter was fixed to pre-outburst value
\end{table*}

\begin{figure}[b] 
\epsscale{1.15} 
\plotone{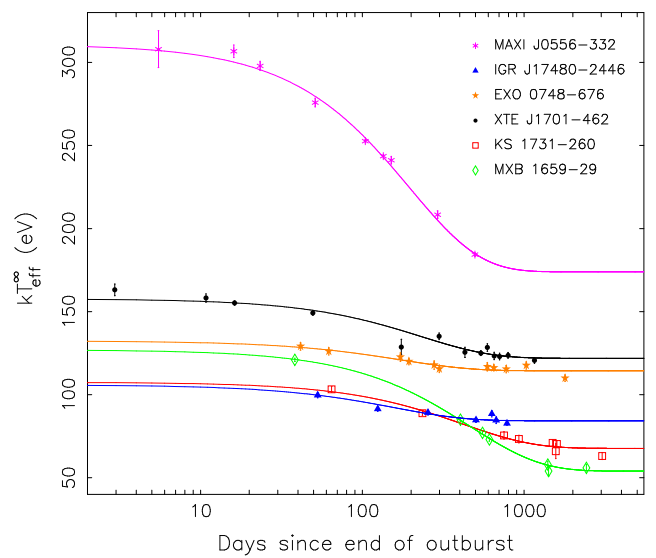} 
\caption{Evolution of the effective temperature of the quiescent neutron star in \maxi, based on fits with model II (purple stars). Temperature data for five other sources are shown as well. The solid lines represent the best fits to the data with an exponential decay to a constant. See Table \ref{tab:expo} for fit parameters and data references.} 
\label{fig:sources}
\end{figure}

The temperatures of the neutron star in \maxi\ during its first $\sim$500 days in quiescence are extraordinarily high compared to those measured for other cooling neutron stars (see Figure \ref{fig:sources}). Even at 500 days into quiescence its temperature is higher than the maximum quiescent temperature  seen in the other five cooling systems. It likely is also  hotter than the neutron star in SAX J1750.8--2900, which was claimed by \citet{lotohe2012} to be the hottest quiescent neutron star in an X-ray binary system, although \maxi\ is expected to continue cooling for a while. The obtained temperatures for \maxi\ are sensitive to the assumed distance (see Table \ref{tab:fits}), but even for distances well below our estimated distance range (e.g., 20 kpc) we find temperatures (134--218 eV for model I and 131--195 eV for model II) that are substantially higher than those observed in \js\ during its first $\sim$500 days (125--163 eV). The short cooling timescale observed in \maxi\ implies a high thermal conductivity of the crust, similar to the other cooling neutron stars that have been studied.

Given the similarities between the outbursts of \maxi\ and \js, it is interesting to compare these two systems in more detail, as it may help us understand what causes the neutron-star crust in \maxi\ to be so hot. \maxi\ was in outburst for $\sim$480 days with a time-averaged luminosity of $\sim$1.7$\times10^{38}(d_{45})^2$ \lum, while \js\ was in outburst for $\sim$585 days with a time-averaged luminosity $\sim$2.0$\times10^{38}(d_{8.8})^2$ \lum\ \citep{frhowi2010}. The total radiated energies of the \maxi\ and \js\ outbursts are therefore 7.1$\times10^{45}(d_{45})^2$ erg and 1.0$\times10^{46}(d_{8.8})^2$ erg, respectively. Despite the fact that the radiated energies and time-averaged luminosities of the two systems are comparable, the initial luminosity of the thermal component (which reflects the temperature at shallow depths in the crust at the end of the outburst)  is an order of magnitude higher  in \maxi\ than in \js.  This suggests the presence of additional shallow heat sources in the crust of \maxi\ and/or that the shallow heat sources in \maxi\ were more efficient per accreted nucleon.

The high observed temperatures are difficult to explain with current crustal heating models. Bringing the initial temperatures down to those seen in \js\ requires a distance of $\sim$10--15 kpc (depending on the assumed model). Such distances are problematic for several reasons. First it implies that Z source behavior in \maxi\ is observed at much lower luminosities (by factors of nine or more) than in other Z sources. Second, fits to the quiescent spectra with such a small distance  are of poor quality. Finally, a smaller distance  does not solve the fact that crustal heating appears to have been much more efficient per accreted nucleon than in other sources. A reduction in distance by a factor of three results in a reduction in luminosity and presumably then, by extension, the total mass accreted onto the neutron star and total heat injected into the crust by a factor of nine. Given that we inferred $\sim$30\% less mass accreted onto the neutron star in \maxi\ during its outburst than in \js\ for our preferred distance of $\sim$45 kpc, this would mean $\sim$12 times less mass accreted onto \maxi\ than \js\, yet similar initial temperatures.

The {\tt nsa} model that we used to fit the thermal emission from the neutron star in \maxi\ did not allow us to explore values of the neutron-star parameters other than $M_{\rm ns}=1.4\,M_\odot$ and $R_{\rm ns}=10$ km, as these parameters are advised to remain fixed  at those values \citep{zapash1996}. While other neutron-star atmosphere models allow for changes in $M_{\rm ns}$ and $R_{\rm ns}$, none of the available models are able to handle the high temperatures observed during the first $\sim$200 days of quiescence. It is, of course, possible that the properties of the neutron star in \maxi\ are significantly different from those in the other cooling neutron-star transients that have been studied. Lower  temperatures would be measured if one assumed a lower $M_{\rm ns}$ and/or a larger $R_{\rm ns}$. To estimate the effects of changes in neutron-star parameters we used the {\tt nsatmos} model to fit the spectrum of observation 11, initially assuming $M_{\rm ns}=1.4\,M_\odot$ and $R_{\rm ns}=$10 km. While keeping the distance from this fit fixed, and changing $M_{\rm ns}$ to 1.2\,$M_\odot$ and $R_{\rm ns}$ to 13 km (values that are still reasonable), the measured temperature was reduced by only $\sim$10\%. Such changes are not large enough to reconcile the temperatures measured in \maxi\ with those of the other sources. 

An alternative explanation for the high inferred temperatures could be that part of the quiescent thermal emission is caused by low-level accretion. Indications for low-level accretion in early quiescence were seen in the form of small flares in the \swift/XRT light curve (Figure \ref{fig:curves}b; see below for a more detailed discussion). The level of accretion in between those flares, where most of our observations were done, is unclear however. \citet{zatuza1995} suggest that low-level accretion onto neutron stars can result in thermal emission, although it would be hard to distinguish it from that of a cooling neutron star \citep{sozaza2011}.  An additional  hard  X-ray  component could be produced by emission from a radiatively inefficient (advection-dominated) accretion flow \citep{namcyi1996} and/or by energetic protons illuminating the neutron-star surface \citep{dedusp2001}. As shown in Section \ref{sec:spectra}, we find that the addition of a power-law to our spectral model (to fit the hard, non-thermal emission) presents a minor, but statistically significant improvement to our fits of the quiescent spectra of \maxi. We also find that, when adding a power-law, the fluxes of the power-law and thermal components are strongly correlated (see Table \ref{tab:bestfit2}), which one would expect if low-level accretion played a significant role.  However, the index of the power-law is unusually low and when fixing it to commonly observed values  ($\Gamma$=1.5--2) the addition of a power-law no longer presents a statistical improvement. It is possible that the {\tt nsa} model simply does not present a good-enough description of the thermal emission from a non-accreting neutron star and that the (shallow) power law is merely fitting the resulting discrepancies; this would likely also result in correlated behavior between the {\tt nsa} and power-law fluxes, as observed.  %{\bf add some text} A comparison with other sources may shed more light on this issue. {\bf rethink}  \citet{cabrde2014} recently studied the decay of an outburst of the neutron-star transient Aql X-1 and it subsequent quiescent phase. They fit the spectra with a combination of a thermal component ({\tt nsatmos}) and a power-law, and find  power-law fractions of 22--40\% and temperatures of 155--170 eV for their quiescent spectra; the decay spectra show higher power-law fractions and temperatures. For comparison, the quiescent spectra of \maxi\ (excluding the two flare observations) have lower power-law fractions (0--35\%, with an average of $\sim$12\%), but much higher temperatures (184--308 eV) than Aql X-1. This suggests that, even if there was a contribution to the thermal emission from accretion in \maxi, its thermal base level, presumably set by the cooling neutron-star crust, is still substantially higher than in other sources.

\citet{pare2013} showed that \js\ likely had not yet reached a steady-state temperature profile at the end of its 19-month outburst. As a result of this, the crust temperature profile had a dip at intermediate densities, leading to a temporary plateau in the temperature decay followed by late-time cooling, both of which have now been observed in \js\ (J.\,K.\ Fridriksson et al., in preparation). Given the outburst similarities between \js\ and \maxi, we expect the cooling curve of \maxi\ to eventually show similar features, although their onset may occur at different times than in \js.

In observations 4 and 6 large increases in the neutron-star temperature and luminosity and power-law luminosity were seen with respect to the overall trend. This behavior is similar to that of two flares seen during the quiescent phase of \js\ \citep{frhowi2010,frhowi2011}. Additional flares can be seen in the \swift\ light curves shown in Figure \ref{fig:curves}b.  Significant flare-like variations in the quiescent luminosity of other NS-LMXBs have been reported as well: e.g., a strong $<$16 d flare in SAX J1750.8--2900 \citep{wide2013}, a similarly strong ($\sim$3$\times10^{34}$ erg\,s$^{-1}$) $<$4 day flare in KS 1741--293 \citep{dewi2013}, and a $\sim$1 week $\sim$9$\times10^{34}$ erg\,s$^{-1}$ flare in GRS J1741--2853 \citep{dewi2009}. Variations on timescales of hours to years have also been observed in the quiescent emission of Cen X-4 \citep{camest1997,caisst2004,cabrmi2010,cabrde2013,becabr2013}, Aql X-1 \citep{rubibr2002a,cafrho2011}, and SAX J1808.4--3658 \citep{castke2008}. The origin of these flares and variations are almost certainly accretion related. The decay timescale of the second flare observed in \js\ is similar to that of the main outburst decay \citep{frhowi2011}, suggesting a common (possibly viscous) timescale. The decay of the first of the flares seen in \maxi\ with \swift, a $\sim$10 d flare which ended just before our first \cha\ observation (see Figure \ref{fig:curves}a), is reasonably well resolved with \swift\ and has a decay timescale of a few days, also similar to that of the outburst decay.  Moreover, the power-law flux increases strongly during the flares and such strongly variable power-law components are not expected from a cooling neutron star, whereas they are commonly associated with low-level accretion, both onto neutron stars and black holes. Finally, the fact that the thermal and non-thermal component are correlated during the flares suggests that the accretion flow is reaching the neutron-star surface.  For this reason, the flare observations are not considered in our modeling of the cooling curves. 

In addition to the early flaring activity, \maxi\ showed a much stronger rebrightening event, starting around day $\sim$170 of quiescence; it lasted for $\sim$60 days, reached a luminosity of $\sim$8$\times10^{37}(d_{45})^2$ erg\,s$^{-1}$, and had an average luminosity of $\sim$4.4$\times10^{37}$ \lum. Although the fluence of this event was a factor of $\sim$33 smaller than that of the main outburst, it is only $\sim$30\% smaller than, e.g., the 2010 outburst of  IGR J17480--2446, in which moderate heating occurred \citep{debrwi2011,dewibr2013}. However, our observations after the large  rebrightening event did not suggest strong deviations from the ongoing cooling trend. 

The distance to \maxi\ remains an  uncertain factor in our fits to the quiescent spectra and modeling of the cooling curve, and this complicates the comparison with other sources. We have employed two methods to estimate the distance to the source: one was a comparison of the CD/HID tracks with those of other Z sources and the other the quiescent spectral fits themselves. In the other cooling neutron-star transients that have been studied, quiescent spectral fits suffer in general from a high degree of degeneracy between the normalization (which depends on the distance and neutron-star radius) and the temperature of the neutron-star atmosphere spectral component. Due to the low effective temperature, the small effective area of the detectors at the lowest energies, and the effects of interstellar absorption, it is primarily the high-energy tail of the thermal component that is being fitted, and usually fits of similar quality can be obtained for a rather wide range of distance (and/or radius) values.  This degeneracy is broken to a significant extent in \maxi\ due to the very low absorption towards the source and the (presumably) much higher temperatures exhibited by it. The fits are quite sensitive to the value of the normalization and it was not possible to achieve fits of adequate quality except for large distance values. Our confidence in such a large value for the distance to the system is strengthened by the fact that the fits with these large distance values also give \nh\ values that are consistent with other measurements, whereas significantly smaller distance values do not, and the fact that the best-fit distance ($\sim$45$\pm$3 kpc for $R_{\rm ns}$ = 10 km) is nicely consistent with the range we estimate from the CD/HID tracks of the source during outburst ($\sim$46$\pm$15 kpc). Our distance estimates  put \maxi\ far into the Galactic halo. While such a location is not common for X-ray binaries, there are several other systems whose properties also suggest very large distances: e.g., GS 1354--64 has a distance greater than 27 kpc \citep{caorzu2009} and IGR J17091--3624 could be as distant as 35 kpc if its mass and luminosity are similar to GRS 1915+105 \citep{albeli2011,wiyaal2012}, although both sources lie close to the Galactic plane. An optical spectrum of the companion star in quiescence might provide a better handle on the distance to \maxi\ than the above methods.

%We note that the drop in count rate by a factor of $\sim$ is much smaller than those seen in other neutron-star transients, such as \js\ and Aql X-1, in which drops by factors of \ldots were found.  Spectral fits?

%Other systems with large distances. Few black holes. \maxi\ not unique in this respect. IGR 1915-like source, 4U 1354, 4U 1957

%\section{Summary}

\acknowledgments

We are grateful to Harvey Tananbaum, Belinda Wilkes, and the {\it Chandra} team for allowing us to use the remaining time of a previously triggered Target of Opportunity program. We also thank Norbert Schartel for approving two \xmm\ Director's Discretionary Time observations, and Neil Gehrels and his team for the flexible scheduling of our \swift\ observations. J.H.\ would like to thank the members of SRON Utrecht, where part of this work was done, for their hospitality. This research has made use of data, software, and/or web tools obtained from NASA's High Energy Astrophysics Science Archive Research Center (HEASARC), a service of Goddard Space Flight Center and the Smithsonian Astrophysical Observatory.
Support for this work was provided by the National Aeronautics and Space Administration through \cha\ Award Number GO2--13052X issued by the \cha\ X-ray Observatory Center, which is operated by the Smithsonian Astrophysical Observatory for and on behalf of the National Aeronautics Space Administration under contract NAS8-03060, and through \swift\ Award Number NNX12AD56G. R.W.\ was supported by an European Research Council starting grant. Finally, we thank the referee for his/her constructive comments.

%\bibliographystyle{aa}
%\bibliography{all-bib}

%\begin{figure}[t] 
%\epsscale{0.7} 
%\plotone{f4_break.eps} 
%\caption{Fits to the cooling data of \maxi.} 
%\label{fig:break}
%\end{figure}

\end{document}